\documentclass[aps,prl,reprint,twocolumn]{revtex4-1}
\usepackage{graphicx}
\usepackage{amsmath}
\usepackage{amssymb}
\usepackage[colorlinks, allcolors=blue]{hyperref}
\usepackage{hypcap}
\usepackage{comment}

\newcommand{\pref}[2]{\hyperref[#1]{\ref{#1}#2}}
\newcommand{\eqpref}[1]{\hyperref[#1]{(\ref{#1})}}

\begin{document}
\title{Ballistic, diffusive, and arrested transport in disordered momentum-space lattices}
\author{Fangzhao Alex An}
\author{Eric J. Meier}
\author{Bryce Gadway}
\email{bgadway@illinois.edu}
\affiliation{Department of Physics, University of Illinois at Urbana-Champaign, Urbana, IL 61801-3080, USA}
\date{\today}
\begin{abstract}
Ultracold atoms in optical lattices offer a unique platform for investigating disorder-driven phenomena.
While static disordered site potentials have been explored in a number of optical lattice experiments, a more general control over site-energy and off-diagonal tunneling disorder has been lacking.
The use of atomic quantum states as ``synthetic dimensions'' has introduced the spectroscopic, site-resolved control necessary to engineer new, more tailored realizations of disorder.
Here, by controlling laser-driven dynamics of atomic population in a momentum-space lattice, we extend the range of synthetic-dimension-based quantum simulation and present the first explorations of dynamical disorder and tunneling disorder in an atomic system.
By applying static tunneling phase disorder to a one-dimensional lattice, we observe ballistic quantum spreading as in the case of uniform tunneling.
When the applied disorder fluctuates on timescales comparable to intersite tunneling, we instead observe diffusive atomic transport, signaling a crossover from quantum to classical expansion dynamics.
We compare these observations to the case of static site-energy disorder, where we directly observe quantum localization in the momentum-space lattice.
\end{abstract}
\maketitle

Over the past two decades, dilute atomic gases have become a fertile testing ground for the study of localization phenomena in disordered quantum systems~\cite{Sanchez-Palencia-Lewenstein-NP-2010}. They have allowed for some of the earliest and most comprehensive studies of Anderson localization of quantum particles~\cite{Moore-Qdeltakicked-1995,Chabe-AndersonMetal-2008,Roati-AndersonLocalization-2008,Billy-AndersonLocalization-2008,KondovAnderson-2011,Jendre-3D-Anderson,Ing-MobEdge}, strongly interacting disordered matter~\cite{Fallani-TowardsBG-2007,White-SpeckleDisorder-2009,Pasienski-Disorder-2010,Gadway-11-PRL,Carrie-Ray-NatPhys,DErrico-1D-Dis}, and many-body localization~\cite{Kondov-MBL,Schreiber842,MBL-Gross,MBL-Vito}.
Still, the emulation of many types of disorder relevant to real systems - e.g., crystal strain and dislocation, site vacancies, interstitial and substitutional defects, magnetic disorder, and thermal phonons - will require new types of control that go beyond traditional methods based on static disorder potentials~\cite{White-SpeckleDisorder-2009}.

\begin{figure}[hb!]
\includegraphics[width=6.75cm]{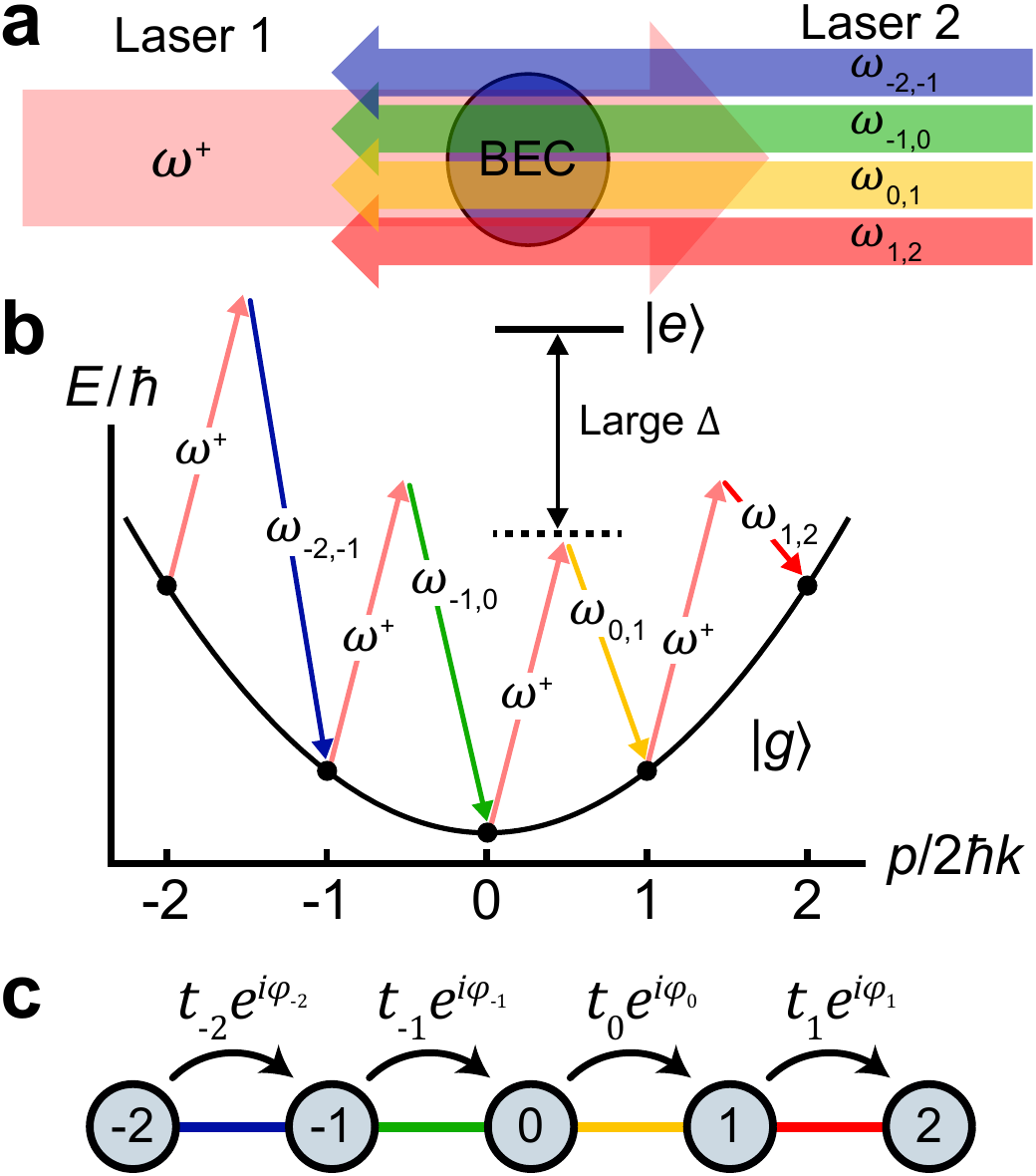}
\caption{\label{FIG:fig1}
\textbf{Spectroscopic control of lattice dynamics.}
\textbf{(a)}~An atomic Bose-Einstein condensate (BEC) illuminated by two counter-propagating lasers, one of which (2) contains multiple discrete spectral components.
\textbf{(b)}~Energy diagram of free-particle-like momentum states coupled by counter-propagating, far-detuned Bragg laser fields (characterized by nearly identical wavevectors $k$). The spectral components of laser 2 are used to separately address individual Bragg transitions.
\textbf{(c)}~Cartoon depiction of the effective tight-binding lattice model when all two-photon Bragg resonance conditions are matched, resulting in a flat site-energy landscape. The amplitudes and phases of the tunneling elements $t_j e^{i\varphi_j}$ are independently controlled through the spectral components $\omega_{j,j+1}$ of laser 2. The lattice site energies $\varepsilon_j$ may also be independently controlled through the detunings from two-photon Bragg resonances.
}
\end{figure}

The recent advent of using atomic quantum states as \emph{synthetic dimensions} has broadened the cold atom toolkit with the spectroscopic, site-resolved control of field-driven transitions~\cite{Celi-ArtificialDim,Stuhl-Edge-2015,Fallani-chiral-2015,Meier-AtomOptics,Meier-SSH,An-FluxLadder}. This technique has aided the study of synthetic gauge fields~\cite{Celi-ArtificialDim,Stuhl-Edge-2015,Fallani-chiral-2015,Wall-Synth,Kolkowitz-Synth,Livi-Synth,An-FluxLadder}, and its spatial and dynamical control offers a prime way to implement specifically tailored, dynamical realizations of disorder that would otherwise be difficult to study. However, current studies based on internal states~\cite{Stuhl-Edge-2015,Fallani-chiral-2015,Wall-Synth,Kolkowitz-Synth,Livi-Synth} have been limited to a small number of sites along the synthetic dimension, inhibiting the study of quantum localization in the presence of disorder.

\begin{figure*}[ht!]
\includegraphics[width=\textwidth]{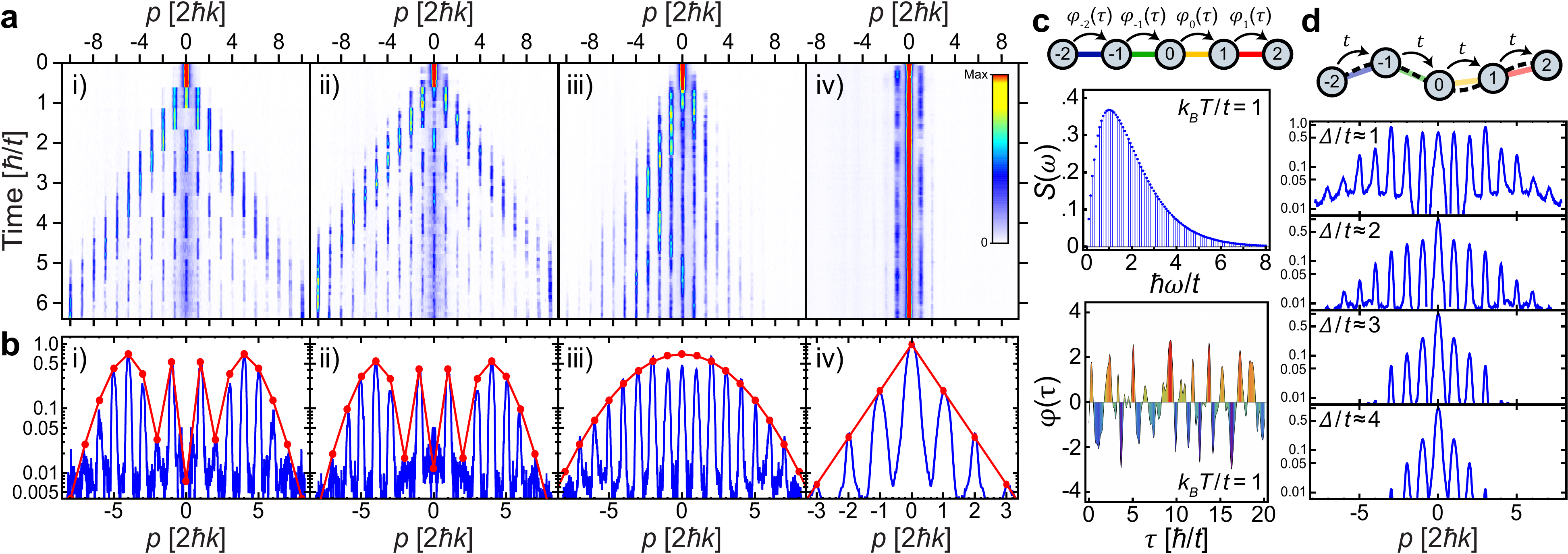}
\caption{\label{FIG:fig2}
	\textbf{Atomic quantum walks in regular and disordered momentum-space lattices.}
	\textbf{(a)}~Nonequilibrium quantum walk dynamics of 1D atomic momentum distributions vs. evolution time for the cases of (i) uniform tunneling, (ii) random static tunneling phases, (iii) random, dynamically-varying tunneling phases characterized by an effective temperature $k_B T / t = 0.66(1)$, and (iv) pseudorandom site energies for $\Delta / t = 5.9(1)$. 
	\textbf{(b)}~Integrated 1D momentum distributions (populations in arbitrary units; symmetrized about zero momentum) after various evolution times $\tau \gtrsim 2.5 \hbar/t$ for the same cases as in (a). For (i) and (ii), we compare to quantum random walk distributions of the form $P_n \propto |J_n (2 \tau t/\hbar)|^2$, for (iii) we compare to a Gaussian distribution $P_n \propto e^{-n^2/2w^2}$, and for (iv) we compare to an exponential distribution $P_n \propto e^{-|n|/\xi}$.
	\textbf{(c)}~Top: Random tunneling phases that vary dynamically with time $\tau$. Middle: Ohmic spectrum of sampled tunneling frequencies for effective temperature $T$, peaked at $\omega = k_BT/\hbar$. Bottom: Representative random tunneling phase dynamics for a specific tunneling element.
	\textbf{(d)}~Top: Pseudorandom site energies, following the form $\varepsilon_n = \Delta \cos(2 \pi b n + \phi)$ of an incommensurate cosine potential (dashed line). Bottom: 1D momentum distributions as in (b,iv) for varying pseudodisorder strengths $\Delta/t$.
}
\end{figure*}

Here, we employ our recently-developed technique of momentum-space lattices~\cite{Gadway-KSPACE,Meier-AtomOptics} to perform the first studies of tailored and dynamical disorder in synthetic dimensions.
Our approach introduces several key advances to cold atom studies of disorder: the achievement of pure off-diagonal tunneling disorder, the dynamical variation of disorder, and site-resolved detection of populations in a disordered system.
For the case of tunneling disorder, we examine the scenario in which only the phase of tunneling is disordered. As expected for a one-dimensional system with only nearest-neighbor tunneling, these random tunneling phases are of zero consequence when applied in a static manner.
When this phase disorder fluctuates on timescales comparable to intersite tunneling, however, we observe a crossover from ballistic to diffusive transport~\cite{LahiniDiffusion}.
We compare to the case of static site-energy disorder, observing Anderson localization at the site-resolved level.

Our bottom-up approach~\cite{Gadway-KSPACE,Meier-AtomOptics} to Hamiltonian engineering is based on the coherent coupling of atomic momentum states to form an effective synthetic lattice of sites in momentum space (see Fig.~\ref{FIG:fig1}).
This approach may be viewed as studying transport in an artificial dimension~\cite{Celi-ArtificialDim} of discrete spatial eigenstates~\cite{NateGold-TrapShake} (as opposed to a bounded set of atomic internal states~\cite{Stuhl-Edge-2015,Fallani-chiral-2015}) through resonant or near-resonant field-driven transitions.

Starting with $^{87}$Rb Bose-Einstein condensates of $\sim$$5 \times 10^4$ atoms, we initiate dynamics between 21 discrete momentum states by applying sets of counter-propagating far-detuned laser fields (wavelength $\lambda = 1064$~nm, wavevector $k = 2\pi/\lambda$), specifically detuned to address multiple two-photon Bragg transitions, as depicted in Fig.~\pref{FIG:fig1}{(a-b)}.
Our spectrally-resolved control of the individual Bragg transitions permits a \emph{local} control of the system parameters, similar to that found in photonic simulators~\cite{Christ-NatRev,And-Light-Seg-07,SzameitReview-2010,AndersonLight-Review,PhotRev-NatPhys-2012}.
Unique to our
implementation
is the direct and arbitrary control of tunneling phases~\cite{Meier-AtomOptics}, and the realized tight-binding model is depicted in Fig.~\pref{FIG:fig1}{(c)}. Here, we use this capability to explore the dynamics of cold atoms subject to disordered and dynamical arrangements of tunneling elements.

Specifically, we explore disorder arising purely in the phase of nearest-neighbor tunneling elements. In higher dimensions, such disordered tunneling phases would give rise to random flux patterns that mimic the physics of charged particles in a random magnetic field~\cite{Lee-Fisher-RandomFlux-1981,RandomField1,RandomField2}. In 1D, however, the absence of closed tunneling paths renders any static arrangement of tunneling phases inconsequential to the dynamical and equilibrium properties of the particle density. Time-varying phases, however, can have a nontrivial influence on the system's dynamical evolution.

We engineer \emph{annealed}, or dynamically varying, disorder of the tunneling phases and study its influence through the atoms' nonequilibrium dynamics following a tunneling quench. Our experiments begin with all population restricted to a single momentum state (site). We suddenly turn on the Bragg laser fields, quenching on the (in general) time-dependent effective Hamiltonian
\begin{equation}
	\hat{H}(\tau) \approx -t \sum_n  (e^{i \varphi_n (\tau)} \hat{c}^\dag_{n+1} \hat{c}_n + \mathrm{h.c.}) + \sum_n \varepsilon_n \hat{c}^\dag_n \hat{c}_n \ ,
	\label{EQ:e0d}
\end{equation}
where $\tau$ is the time variable, $t$ is the (homogeneous) tunneling energy, and $\hat{c}_n$ ($\hat{c}^\dagger_n$) is the annihilation (creation) operator for the momentum state with index $n$ (momentum $p_n = 2n\hbar k$). The tunneling phases $\varphi_n$ and site energies $\varepsilon_n$ are controlled through the phases and detunings of the two-photon momentum Bragg transitions, respectively. After a variable duration of laser-driven dynamics, we perform direct absorption imaging of the final distribution of momentum states, which naturally separate during 18~ms time of flight. Analysis of these distributions, including determination of site populations through a multi-Gaussian fit, is as described in Ref.~\cite{Meier-AtomOptics}.

As a control, we first examine the case of no disorder, with all site-energies set to zero and uniform, static tunneling phases $\varphi_n(\tau) = \varphi$. Figure ~\pref{FIG:fig2}{(a,i)} shows the evolution of the 1D momentum distribution, obtained from time-of-flight images integrated along the axis normal to the imparted momentum, displaying ballistic expansion characteristic of a continuous-time quantum walk. For times before the atoms reflect from the open boundaries of the 21-site lattice, we find good qualitative agreement between the observed momentum distributions and the expected form $P_n = |J_n (\vartheta)|^2$, where $J_n$ is the Bessel function of order $n$ and $\vartheta = 2 \tau t/\hbar$. Figure~\pref{FIG:fig2}{(b,i)} shows the (symmetrized) momentum profile at time $\tau \sim 3 \hbar/t$ along with the Bessel function distribution for $\vartheta = 5.4$.

In comparison, Fig.~\pref{FIG:fig2}{(a,ii)} shows the case of zero site energies and static, random tunneling phases $\varphi_n \in [0,2\pi)$. The dynamics are nearly identical to the case of uniform tunneling phases. This is consistent with the expectation that any pattern of static tunneling phases in 1D is irrelevant for the dynamics of the effective tight-binding model realized by our controlled laser coupling. For this case, Fig.~\pref{FIG:fig2}{(b,ii)} shows the (symmetrized) momentum profile at $\tau \sim 2.5 \hbar/t$ along with the Bessel function distribution for $\vartheta = 5.35$.

While static phase disorder has little impact on the quantum random walk dynamics, we may generally expect that controlled random phase jumps or even pseudorandom variations of the phases should inhibit coherent transport, mimicking random phase shifts induced through interaction with a thermal environment.
To probe such behavior, we implement dynamical phase disorder by composing each tunneling phase $\varphi_n$ from a broad spectrum of oscillatory terms with randomly-defined phases $\theta_{n,i}$ but well-defined frequencies $\omega_i$, the weights of which are derived from an ohmic bath distribution. Specifically, the dynamical tunneling phases take the form
\begin{equation*}
\varphi_n(\tau) = 4\pi \sum_{i = 1}^N S(\omega_i) \cos (\omega_i \tau + \theta_{n,i}) / \sum_{i = 1}^N S(\omega_i),
\end{equation*}
where $S(\omega) = (\hbar \omega / k_B T) \mathrm{exp}[-(\hbar \omega / k_B T)]$,
the $\theta_{n,i}$ are randomly chosen from $[0, 2\pi )$, and $T$ is an artificial temperature scale that sets the range of the frequency distribution. In this discrete formulation of $\varphi_n(\tau)$, we include $N=50$ frequencies ranging between zero and $8 k_B T / \hbar$. The frequency spectrum and dynamics for one tunneling phase $\varphi_n(\tau)$ are shown in Fig.~\pref{FIG:fig2}{(c)} for the case of $k_BT/t = 1$.

Figure~\pref{FIG:fig2}{(a,iii)} displays the population dynamics in the presence of this dynamical disorder, characterized by an effective temperature $k_B T/t = 0.66(1)$ and averaged over three independent realizations of the disorder. The dynamics no longer feature ballistically separating wavepackets, instead displaying a broad, slowly spreading distribution peaked near zero momentum. A clear deviation of the (symmetrized) momentum distribution from the form $P_n = |J_n (\vartheta)|^2$ describing the previous quantum walk dynamics can be seen in Fig.~\pref{FIG:fig2}{(b,iii)} (shown at the time $\tau \sim 3.8 \hbar/t$). The displayed Gaussian population distribution gives much better agreement, consistent with spreading governed by an effectively classical or thermal random walk.

Lastly, while no influence of static tunneling phase disorder is expected in 1D, the effect of static site-energy disorder is dramatically different.
Here, with homogeneous static tunneling terms, we explore the influence of pseudorandom variations of the site energies governed by the Aubry-Andr\'{e} model~\cite{Fallani-TowardsBG-2007,Roati-AndersonLocalization-2008,Gadway-11-PRL,Schreiber842}.
With an irrational periodicity $b=(\sqrt{5}-1)/2$, the site energies $\varepsilon_n = \Delta \cos(2 \pi b n + \phi)$ do not repeat, and are governed by a pseudorandom distribution.
For an infinite system, this Aubry-Andr\'{e} model with diagonal disorder features a metal-insulator transition at the critical disorder strength $\Delta_c = 2 t$.
The expansion dynamics for the strong disorder case $\Delta / t = 5.9(1)$ are shown in Fig.~\pref{FIG:fig2}{(a,iv)}, with population largely restricted to the initial, central momentum order. The exponentially localized distribution of site populations (symmetrized and averaged over all profiles in the range $\tau \sim 5-6.3 \hbar/t$) is shown in Fig.~\pref{FIG:fig2}{(b,iv)}, along with an exponential distribution with localization length $\xi = 0.6$ lattice sites. Analogous population distributions (again symmetrized and averaged over the same time range) are shown for the cases of weaker disorder [$\Delta/t = 0.98(1), 1.96(3), 3.05(4), 4.02(9)$] in Fig.~\pref{FIG:fig2}{(d)}, exhibiting an apparent transition to exponential localization for $\Delta/t \gtrsim 2$.

\begin{figure*}[t!]
\includegraphics[width=\textwidth]{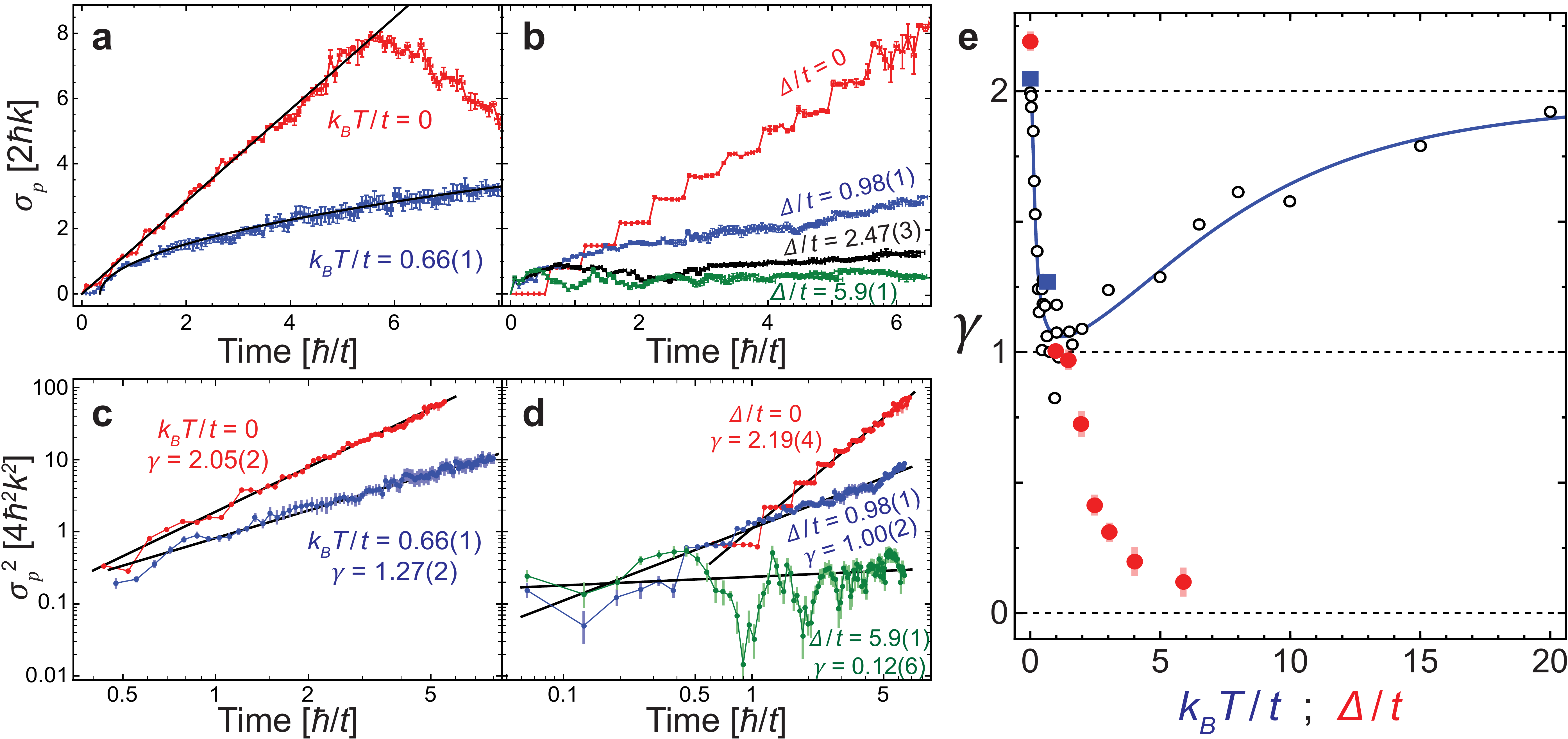}
\caption{\label{FIG:fig3}
	\textbf{Expansion dynamics in static and dynamical disorder.}
	\textbf{(a)}~Momentum width $\sigma_p$ (standard deviation, units of $2 \hbar k$) vs. evolution time ($\tau$, units of $\hbar/t$) for random static tunneling phases (red data, labeled $k_BT/t = 0$) and random dynamical tunneling phases (blue data, labeled $k_BT/t = 0.66(1)$). Overlaid as black lines are the predicted dynamics for ballistic ($\sigma_p = \sqrt{2} \tau$) and diffusive transport ($\sigma_p = \sqrt{2 \tau}$, shifted by $0.35 \, \hbar /t$).
	\textbf{(b)}~Momentum-width dynamics for the cases of static site-energy pseudodisorder and uniform equal-phase tunneling. The data curves relate to disorder strengths of $\Delta/t = 0$ (red data), $\Delta/t = 0.98(1)$ (blue data), $\Delta/t = 2.47(3)$ (black data), and $\Delta/t = 5.9(1)$ (green data).
	\textbf{(c)}~Double logarithmic plot of the momentum variance ($\sigma_p^2$, in units of $4 \hbar^2 k^2$) for the random phase data in (a), fit to the form $V(\tau) = \alpha \tau^{\gamma}$. The fit-determined values of $\gamma$ are shown for each case.
	\textbf{(d)}~Double logarithmic plot of the momentum variance for the static disorder data in (c), along with power-law fits and extracted expansion exponents $\gamma$.
	\textbf{(e)}~The fit-determined expansion exponents $\gamma$ plotted versus the effective annealed disorder temperature ($k_BT/t$, blue squares) for dynamical disorder and versus the disorder strength ($\Delta/t$, red circles) for static pseudodisorder. The solid blue line is a fit to numerical simulations (open black circles) for the case of dynamically varying phase disorder.
}
\end{figure*}

For all of the explored cases, we study these expansion dynamics in greater detail in Fig.~\ref{FIG:fig3}.
Figure~\pref{FIG:fig3}{(a)} examines the momentum-width ($\sigma_p$) dynamics of the atomic distributions for the cases of static and dynamic random phase disorder. For static phase disorder, we observe a roughly linear increase of $\sigma_p$ until population reflects from the open system boundaries, while dynamical phase disorder leads to sub-ballistic expansion.
In particular, for time $\tau$ measured in units of $\hbar/t$ and momentum-width $\sigma_p$ in units of the site separation $2 \hbar k$, these two cases agree well with the displayed theory curves for ballistic and diffusive expansion, having the forms $\sigma_p = \sqrt{2} \tau$ and $\sigma_p = \sqrt{2 \tau}$, respectively (with the latter curve shifted by $0.35 \hbar /t$).
To explore these two different expansions more quantitatively, we fit the momentum variance $ V_p \equiv \sigma_p^2$ to a power-law $V_p(\tau) = \alpha \tau^{\gamma}$, performing a linear fit to variance dynamics on a double logarithmic scale as shown in Fig.~\pref{FIG:fig3}{(c)}.
The fit-determined expansion exponents $\gamma$ for the cases of static and dynamically disordered tunneling phases are $2.05(2)$ and $1.27(2)$, respectively. These values are roughly consistent with a coherent, quantum random walk for the case of static tunneling phases and an incoherent, nearly diffusive random walk for the case of dynamical phase disorder.

The observed transport dynamics cross over from ballistic to diffusive as the effective thermal energy scale $k_B T$ approaches the coherent tunneling energy $t$, matching our expectation that randomly-varying tunneling phases can mimic the random dephasing induced by a thermal environment.
We note that similar classical random walk behavior has been seen previously for both atoms and photons, due to irreversible decoherence~\cite{QtoC-Theory-Decoh,Broome-White-QtoC-Decoherence,Silberhorn-DisorderAndDecoherence,Karski174} and thermal excitations~\cite{fukuhara:quantum_2013}.
However, this is the first observation based on reversible engineered ``noise'' of a Hamiltonian parameter.
These observations of a thermal random walk suggest that annealed disorder may provide a means of mimicking thermal fluctuations and studying thermodynamical properties \cite{Osterloh-NonAbel-2005PRL} of simulated models using atomic momentum-space lattices, and by extension other nonequilibrium experimental platforms such as photonic simulators.

We also analyze the full expansion dynamics for the case of static site energy disorder in Figs.~\pref{FIG:fig3}{(b,d)}. For homogeneous static tunnelings and thus zero disorder ($\Delta/t =0$), we observe momentum-width dynamics similar to the case of static random tunneling phases, but with one distinct difference: while $\sigma_p$ features a linear increase for random static phases, it increases in a step-wise fashion for uniform tunneling phases. This slight disagreement is a byproduct of the underlying laser-driven dynamics that give rise to the effective tight-binding model described by Eq.~\ref{EQ:e0d}. The Bragg laser field 2 (see Fig.~\ref{FIG:fig1}) features a comb of 20 discrete, equally-spaced frequencies, each of which primarily addresses a single Bragg transition. Weak off-resonant coupling terms conspire to produce this step-like behavior in the case of equal-phase driving, while this behavior is mostly absent for random tunneling phases.

Evolution of the momentum-width ($\sigma_p$) for the site-energy disorder cases of $\Delta/t = 0.98(1), 2.47(3), 5.9(1)$ are also shown in Fig.~\pref{FIG:fig3}{(b)}. We observe the reduction of expansion dynamics with increasing disorder, with nearly arrested dynamics in the strong disorder limit. More quantitatively, fits of the variance dynamics as shown in Fig.~\pref{FIG:fig3}{(d)} reveal sub-ballistic, nearly diffusive expansion for intermediate disorder [$\gamma = 1.00(2)$ for $\Delta/t = 0.98(1)$], giving way to a nearly vanishing expansion exponent for strong disorder [$\gamma = 0.12(6)$ for $\Delta/t = 5.9(1)$].

The extracted expansion exponents for all of the explored cases are summarized in Fig.~\pref{FIG:fig3}{(e)}. For static site-energy disorder (red circles), while longer expansion times than those explored ($\tau \lesssim 6.3 \hbar/t$) would better distinguish insulating behavior from sub-ballistic and sub-diffusive expansion, a clear trend towards arrested transport ($\gamma \sim 0$) is found for $\Delta/t \gg 1$. Combined with the observation of exponential localization of the site populations in Fig.~\pref{FIG:fig2}{(b,iv)} and Fig.~\pref{FIG:fig2}{(d)}, these observations are consistent with a crossover in our 21-site system from metallic behavior to quantum localization for $\Delta/t \gtrsim 2$.

Our observations of a crossover from ballistic expansion ($\gamma \sim 2$) to nearly diffusive transport ($\gamma \sim 1$) for randomly fluctuating tunneling phase disorder are also summarized in Fig.~\pref{FIG:fig3}{(e)}.
In the experimentally-accessible regime of low to moderate effective thermal energies ($k_B T / t \lesssim 1$), our experimental data points (blue squares) match up well with numerical simulation (open black circles).
For the magnitude of tunneling energy used in these experiments, we are restricted from exploring higher effective temperatures ($k_B T /t \gtrsim 1$), as rapid variations of the tunneling phases introduce spurious spectral components of the Bragg laser fields that could drive undesired transitions.
Simulations in this high-temperature regime suggest that the expansion exponent should rise back up for increasing temperatures, saturating to a value $\gamma \sim 2$.
This results from the fact that the time-averaged phase effectively vanishes when the timescale of pseudorandom phase variations is much shorter than the tunneling time.

The demonstrated levels of local and time-dependent control over tunneling elements and site energies in our synthetic momentum-space lattice have allowed us to perform first-of-their-kind explorations of annealed disorder in an atomic system. Such an approach based on synthetic dimensions should enable myriad future explorations of engineered Floquet dynamics~\cite{Rudner,Goldman-Floq-And,Szameit-Floq-And,Titum-Floq-And} and novel disordered lattices~\cite{Kosior-RandomFractals,Dunlap-1990}.
Furthermore, the realization of designer disorder in a system that supports nonlinear atomic interactions~\cite{Rolston-NL-2002} should permit us to explore novel aspects of many-body localization~\cite{aleiner:finite_temperature_disorder_2010}.


\begin{thebibliography}{52}%
\makeatletter
\providecommand \@ifxundefined [1]{%
 \@ifx{#1\undefined}
}%
\providecommand \@ifnum [1]{%
 \ifnum #1\expandafter \@firstoftwo
 \else \expandafter \@secondoftwo
 \fi
}%
\providecommand \@ifx [1]{%
 \ifx #1\expandafter \@firstoftwo
 \else \expandafter \@secondoftwo
 \fi
}%
\providecommand \natexlab [1]{#1}%
\providecommand \enquote  [1]{``#1''}%
\providecommand \bibnamefont  [1]{#1}%
\providecommand \bibfnamefont [1]{#1}%
\providecommand \citenamefont [1]{#1}%
\providecommand \href@noop [0]{\@secondoftwo}%
\providecommand \href [0]{\begingroup \@sanitize@url \@href}%
\providecommand \@href[1]{\@@startlink{#1}\@@href}%
\providecommand \@@href[1]{\endgroup#1\@@endlink}%
\providecommand \@sanitize@url [0]{\catcode `\\12\catcode `\$12\catcode
  `\&12\catcode `\#12\catcode `\^12\catcode `\_12\catcode `\%12\relax}%
\providecommand \@@startlink[1]{}%
\providecommand \@@endlink[0]{}%
\providecommand \url  [0]{\begingroup\@sanitize@url \@url }%
\providecommand \@url [1]{\endgroup\@href {#1}{\urlprefix }}%
\providecommand \urlprefix  [0]{URL }%
\providecommand \Eprint [0]{\href }%
\providecommand \doibase [0]{http://dx.doi.org/}%
\providecommand \selectlanguage [0]{\@gobble}%
\providecommand \bibinfo  [0]{\@secondoftwo}%
\providecommand \bibfield  [0]{\@secondoftwo}%
\providecommand \translation [1]{[#1]}%
\providecommand \BibitemOpen [0]{}%
\providecommand \bibitemStop [0]{}%
\providecommand \bibitemNoStop [0]{.\EOS\space}%
\providecommand \EOS [0]{\spacefactor3000\relax}%
\providecommand \BibitemShut  [1]{\csname bibitem#1\endcsname}%
\let\auto@bib@innerbib\@empty
\bibitem [{\citenamefont {Sanchez-Palencia}\ and\ \citenamefont
  {Lewenstein}(2010)}]{Sanchez-Palencia-Lewenstein-NP-2010}%
  \BibitemOpen
  \bibfield  {author} {\bibinfo {author} {\bibfnamefont {L.}~\bibnamefont
  {Sanchez-Palencia}}\ and\ \bibinfo {author} {\bibfnamefont {M.}~\bibnamefont
  {Lewenstein}},\ }\href {\doibase 10.1038/nphys1507} {\bibfield  {journal}
  {\bibinfo  {journal} {Nat. Phys.}\ }\textbf {\bibinfo {volume} {6}},\
  \bibinfo {pages} {87} (\bibinfo {year} {2010})}\BibitemShut {NoStop}%
\bibitem [{\citenamefont {Moore}\ \emph {et~al.}(1995)\citenamefont {Moore},
  \citenamefont {Robinson}, \citenamefont {Bharucha}, \citenamefont
  {Sundaram},\ and\ \citenamefont {Raizen}}]{Moore-Qdeltakicked-1995}%
  \BibitemOpen
  \bibfield  {author} {\bibinfo {author} {\bibfnamefont {F.~L.}\ \bibnamefont
  {Moore}}, \bibinfo {author} {\bibfnamefont {J.~C.}\ \bibnamefont {Robinson}},
  \bibinfo {author} {\bibfnamefont {C.~F.}\ \bibnamefont {Bharucha}}, \bibinfo
  {author} {\bibfnamefont {B.}~\bibnamefont {Sundaram}}, \ and\ \bibinfo
  {author} {\bibfnamefont {M.~G.}\ \bibnamefont {Raizen}},\ }\href {\doibase
  10.1103/PhysRevLett.75.4598} {\bibfield  {journal} {\bibinfo  {journal}
  {Phys. Rev. Lett.}\ }\textbf {\bibinfo {volume} {75}},\ \bibinfo {pages}
  {4598} (\bibinfo {year} {1995})}\BibitemShut {NoStop}%
\bibitem [{\citenamefont {Chab\'{e}}\ \emph {et~al.}(2008)\citenamefont
  {Chab\'{e}}, \citenamefont {Lemari\'{e}}, \citenamefont {Gr\'{e}maud},
  \citenamefont {Delande}, \citenamefont {Szriftgiser},\ and\ \citenamefont
  {Garreau}}]{Chabe-AndersonMetal-2008}%
  \BibitemOpen
  \bibfield  {author} {\bibinfo {author} {\bibfnamefont {J.}~\bibnamefont
  {Chab\'{e}}}, \bibinfo {author} {\bibfnamefont {G.}~\bibnamefont
  {Lemari\'{e}}}, \bibinfo {author} {\bibfnamefont {B.}~\bibnamefont
  {Gr\'{e}maud}}, \bibinfo {author} {\bibfnamefont {D.}~\bibnamefont
  {Delande}}, \bibinfo {author} {\bibfnamefont {P.}~\bibnamefont
  {Szriftgiser}}, \ and\ \bibinfo {author} {\bibfnamefont {J.~C.}\ \bibnamefont
  {Garreau}},\ }\href {\doibase 10.1103/PhysRevLett.101.255702} {\bibfield
  {journal} {\bibinfo  {journal} {Phys. Rev. Lett.}\ }\textbf {\bibinfo
  {volume} {101}},\ \bibinfo {pages} {255702} (\bibinfo {year}
  {2008})}\BibitemShut {NoStop}%
\bibitem [{\citenamefont {Roati}\ \emph {et~al.}(2008)\citenamefont {Roati},
  \citenamefont {D'Errico}, \citenamefont {Fallani}, \citenamefont {Fattori},
  \citenamefont {Fort}, \citenamefont {Zaccanti}, \citenamefont {Modugno},
  \citenamefont {Modugno},\ and\ \citenamefont
  {Inguscio}}]{Roati-AndersonLocalization-2008}%
  \BibitemOpen
  \bibfield  {author} {\bibinfo {author} {\bibfnamefont {G.}~\bibnamefont
  {Roati}}, \bibinfo {author} {\bibfnamefont {C.}~\bibnamefont {D'Errico}},
  \bibinfo {author} {\bibfnamefont {L.}~\bibnamefont {Fallani}}, \bibinfo
  {author} {\bibfnamefont {M.}~\bibnamefont {Fattori}}, \bibinfo {author}
  {\bibfnamefont {C.}~\bibnamefont {Fort}}, \bibinfo {author} {\bibfnamefont
  {M.}~\bibnamefont {Zaccanti}}, \bibinfo {author} {\bibfnamefont
  {G.}~\bibnamefont {Modugno}}, \bibinfo {author} {\bibfnamefont
  {M.}~\bibnamefont {Modugno}}, \ and\ \bibinfo {author} {\bibfnamefont
  {M.}~\bibnamefont {Inguscio}},\ }\href {\doibase 10.1038/nature07071}
  {\bibfield  {journal} {\bibinfo  {journal} {Nature}\ }\textbf {\bibinfo
  {volume} {453}},\ \bibinfo {pages} {895} (\bibinfo {year}
  {2008})}\BibitemShut {NoStop}%
\bibitem [{\citenamefont {Billy}\ \emph {et~al.}(2008)\citenamefont {Billy},
  \citenamefont {Josse}, \citenamefont {Zuo}, \citenamefont {Bernard},
  \citenamefont {Hambrecht}, \citenamefont {Lugan}, \citenamefont
  {Cl\'{e}ment}, \citenamefont {Sanchez-Palencia}, \citenamefont {Bouyer},\
  and\ \citenamefont {Aspect}}]{Billy-AndersonLocalization-2008}%
  \BibitemOpen
  \bibfield  {author} {\bibinfo {author} {\bibfnamefont {J.}~\bibnamefont
  {Billy}}, \bibinfo {author} {\bibfnamefont {V.}~\bibnamefont {Josse}},
  \bibinfo {author} {\bibfnamefont {Z.}~\bibnamefont {Zuo}}, \bibinfo {author}
  {\bibfnamefont {A.}~\bibnamefont {Bernard}}, \bibinfo {author} {\bibfnamefont
  {B.}~\bibnamefont {Hambrecht}}, \bibinfo {author} {\bibfnamefont
  {P.}~\bibnamefont {Lugan}}, \bibinfo {author} {\bibfnamefont
  {D.}~\bibnamefont {Cl\'{e}ment}}, \bibinfo {author} {\bibfnamefont
  {L.}~\bibnamefont {Sanchez-Palencia}}, \bibinfo {author} {\bibfnamefont
  {P.}~\bibnamefont {Bouyer}}, \ and\ \bibinfo {author} {\bibfnamefont
  {A.}~\bibnamefont {Aspect}},\ }\href {\doibase 10.1038/nature07000}
  {\bibfield  {journal} {\bibinfo  {journal} {Nature}\ }\textbf {\bibinfo
  {volume} {453}},\ \bibinfo {pages} {891} (\bibinfo {year}
  {2008})}\BibitemShut {NoStop}%
\bibitem [{\citenamefont {Kondov}\ \emph {et~al.}(2011)\citenamefont {Kondov},
  \citenamefont {McGehee}, \citenamefont {Zirbel},\ and\ \citenamefont
  {DeMarco}}]{KondovAnderson-2011}%
  \BibitemOpen
  \bibfield  {author} {\bibinfo {author} {\bibfnamefont {S.~S.}\ \bibnamefont
  {Kondov}}, \bibinfo {author} {\bibfnamefont {W.~R.}\ \bibnamefont {McGehee}},
  \bibinfo {author} {\bibfnamefont {J.~J.}\ \bibnamefont {Zirbel}}, \ and\
  \bibinfo {author} {\bibfnamefont {B.}~\bibnamefont {DeMarco}},\ }\href
  {\doibase 10.1126/science.1209019} {\bibfield  {journal} {\bibinfo  {journal}
  {Science}\ }\textbf {\bibinfo {volume} {334}},\ \bibinfo {pages} {66}
  (\bibinfo {year} {2011})}\BibitemShut {NoStop}%
\bibitem [{\citenamefont {Jendrzejewski}\ \emph {et~al.}(2012)\citenamefont
  {Jendrzejewski}, \citenamefont {Bernard}, \citenamefont {M\"{u}ller},
  \citenamefont {Cheinet}, \citenamefont {Josse}, \citenamefont {Piraud},
  \citenamefont {Pezz\'{e}}, \citenamefont {Sanchez-Palencia}, \citenamefont
  {Aspect},\ and\ \citenamefont {Bouyer}}]{Jendre-3D-Anderson}%
  \BibitemOpen
  \bibfield  {author} {\bibinfo {author} {\bibfnamefont {F.}~\bibnamefont
  {Jendrzejewski}}, \bibinfo {author} {\bibfnamefont {A.}~\bibnamefont
  {Bernard}}, \bibinfo {author} {\bibfnamefont {K.}~\bibnamefont {M\"{u}ller}},
  \bibinfo {author} {\bibfnamefont {P.}~\bibnamefont {Cheinet}}, \bibinfo
  {author} {\bibfnamefont {V.}~\bibnamefont {Josse}}, \bibinfo {author}
  {\bibfnamefont {M.}~\bibnamefont {Piraud}}, \bibinfo {author} {\bibfnamefont
  {L.}~\bibnamefont {Pezz\'{e}}}, \bibinfo {author} {\bibfnamefont
  {L.}~\bibnamefont {Sanchez-Palencia}}, \bibinfo {author} {\bibfnamefont
  {A.}~\bibnamefont {Aspect}}, \ and\ \bibinfo {author} {\bibfnamefont
  {P.}~\bibnamefont {Bouyer}},\ }\href {\doibase 10.1038/nphys2256} {\bibfield
  {journal} {\bibinfo  {journal} {Nat. Phys.}\ }\textbf {\bibinfo {volume}
  {8}},\ \bibinfo {pages} {398} (\bibinfo {year} {2012})}\BibitemShut {NoStop}%
\bibitem [{\citenamefont {Semeghini}\ \emph {et~al.}(2015)\citenamefont
  {Semeghini}, \citenamefont {Landini}, \citenamefont {Castilho}, \citenamefont
  {Roy}, \citenamefont {Spagnolli}, \citenamefont {Trenkwalder}, \citenamefont
  {Fattori}, \citenamefont {Inguscio},\ and\ \citenamefont
  {Modugno}}]{Ing-MobEdge}%
  \BibitemOpen
  \bibfield  {author} {\bibinfo {author} {\bibfnamefont {G.}~\bibnamefont
  {Semeghini}}, \bibinfo {author} {\bibfnamefont {M.}~\bibnamefont {Landini}},
  \bibinfo {author} {\bibfnamefont {P.}~\bibnamefont {Castilho}}, \bibinfo
  {author} {\bibfnamefont {S.}~\bibnamefont {Roy}}, \bibinfo {author}
  {\bibfnamefont {G.}~\bibnamefont {Spagnolli}}, \bibinfo {author}
  {\bibfnamefont {A.}~\bibnamefont {Trenkwalder}}, \bibinfo {author}
  {\bibfnamefont {M.}~\bibnamefont {Fattori}}, \bibinfo {author} {\bibfnamefont
  {M.}~\bibnamefont {Inguscio}}, \ and\ \bibinfo {author} {\bibfnamefont
  {G.}~\bibnamefont {Modugno}},\ }\href {\doibase 10.1038/nphys3339} {\bibfield
   {journal} {\bibinfo  {journal} {Nat. Phys.}\ }\textbf {\bibinfo {volume}
  {11}},\ \bibinfo {pages} {554} (\bibinfo {year} {2015})}\BibitemShut
  {NoStop}%
\bibitem [{\citenamefont {Fallani}\ \emph {et~al.}(2007)\citenamefont
  {Fallani}, \citenamefont {Lye}, \citenamefont {Guarrera}, \citenamefont
  {Fort},\ and\ \citenamefont {Inguscio}}]{Fallani-TowardsBG-2007}%
  \BibitemOpen
  \bibfield  {author} {\bibinfo {author} {\bibfnamefont {L.}~\bibnamefont
  {Fallani}}, \bibinfo {author} {\bibfnamefont {J.~E.}\ \bibnamefont {Lye}},
  \bibinfo {author} {\bibfnamefont {V.}~\bibnamefont {Guarrera}}, \bibinfo
  {author} {\bibfnamefont {C.}~\bibnamefont {Fort}}, \ and\ \bibinfo {author}
  {\bibfnamefont {M.}~\bibnamefont {Inguscio}},\ }\href {\doibase
  10.1103/PhysRevLett.98.130404} {\bibfield  {journal} {\bibinfo  {journal}
  {Phys. Rev. Lett.}\ }\textbf {\bibinfo {volume} {98}},\ \bibinfo {pages}
  {130404} (\bibinfo {year} {2007})}\BibitemShut {NoStop}%
\bibitem [{\citenamefont {White}\ \emph {et~al.}(2009)\citenamefont {White},
  \citenamefont {Pasienski}, \citenamefont {McKay}, \citenamefont {Zhou},
  \citenamefont {Ceperley},\ and\ \citenamefont
  {DeMarco}}]{White-SpeckleDisorder-2009}%
  \BibitemOpen
  \bibfield  {author} {\bibinfo {author} {\bibfnamefont {M.}~\bibnamefont
  {White}}, \bibinfo {author} {\bibfnamefont {M.}~\bibnamefont {Pasienski}},
  \bibinfo {author} {\bibfnamefont {D.}~\bibnamefont {McKay}}, \bibinfo
  {author} {\bibfnamefont {S.~Q.}\ \bibnamefont {Zhou}}, \bibinfo {author}
  {\bibfnamefont {D.}~\bibnamefont {Ceperley}}, \ and\ \bibinfo {author}
  {\bibfnamefont {B.}~\bibnamefont {DeMarco}},\ }\href {\doibase
  10.1103/PhysRevLett.102.055301} {\bibfield  {journal} {\bibinfo  {journal}
  {Phys. Rev. Lett.}\ }\textbf {\bibinfo {volume} {102}},\ \bibinfo {pages}
  {055301} (\bibinfo {year} {2009})}\BibitemShut {NoStop}%
\bibitem [{\citenamefont {Pasienski}\ \emph {et~al.}(2010)\citenamefont
  {Pasienski}, \citenamefont {McKay}, \citenamefont {White},\ and\
  \citenamefont {DeMarco}}]{Pasienski-Disorder-2010}%
  \BibitemOpen
  \bibfield  {author} {\bibinfo {author} {\bibfnamefont {M.}~\bibnamefont
  {Pasienski}}, \bibinfo {author} {\bibfnamefont {D.}~\bibnamefont {McKay}},
  \bibinfo {author} {\bibfnamefont {M.}~\bibnamefont {White}}, \ and\ \bibinfo
  {author} {\bibfnamefont {B.}~\bibnamefont {DeMarco}},\ }\href {\doibase
  10.1038/nphys1726} {\bibfield  {journal} {\bibinfo  {journal} {Nat. Phys.}\
  }\textbf {\bibinfo {volume} {6}},\ \bibinfo {pages} {677} (\bibinfo {year}
  {2010})}\BibitemShut {NoStop}%
\bibitem [{\citenamefont {Gadway}\ \emph {et~al.}(2011)\citenamefont {Gadway},
  \citenamefont {Pertot}, \citenamefont {Reeves}, \citenamefont {Vogt},\ and\
  \citenamefont {Schneble}}]{Gadway-11-PRL}%
  \BibitemOpen
  \bibfield  {author} {\bibinfo {author} {\bibfnamefont {B.}~\bibnamefont
  {Gadway}}, \bibinfo {author} {\bibfnamefont {D.}~\bibnamefont {Pertot}},
  \bibinfo {author} {\bibfnamefont {J.}~\bibnamefont {Reeves}}, \bibinfo
  {author} {\bibfnamefont {M.}~\bibnamefont {Vogt}}, \ and\ \bibinfo {author}
  {\bibfnamefont {D.}~\bibnamefont {Schneble}},\ }\href {\doibase
  10.1103/PhysRevLett.107.145306} {\bibfield  {journal} {\bibinfo  {journal}
  {Phys. Rev. Lett.}\ }\textbf {\bibinfo {volume} {107}},\ \bibinfo {pages}
  {145306} (\bibinfo {year} {2011})}\BibitemShut {NoStop}%
\bibitem [{\citenamefont {Meldgin}\ \emph {et~al.}(2016)\citenamefont
  {Meldgin}, \citenamefont {Ray}, \citenamefont {Russ}, \citenamefont {Chen},
  \citenamefont {Ceperley},\ and\ \citenamefont
  {DeMarco}}]{Carrie-Ray-NatPhys}%
  \BibitemOpen
  \bibfield  {author} {\bibinfo {author} {\bibfnamefont {C.}~\bibnamefont
  {Meldgin}}, \bibinfo {author} {\bibfnamefont {U.}~\bibnamefont {Ray}},
  \bibinfo {author} {\bibfnamefont {P.}~\bibnamefont {Russ}}, \bibinfo {author}
  {\bibfnamefont {D.}~\bibnamefont {Chen}}, \bibinfo {author} {\bibfnamefont
  {D.~M.}\ \bibnamefont {Ceperley}}, \ and\ \bibinfo {author} {\bibfnamefont
  {B.}~\bibnamefont {DeMarco}},\ }\href {\doibase 10.1038/nphys3695} {\bibfield
   {journal} {\bibinfo  {journal} {Nat. Phys.}\ }\textbf {\bibinfo {volume}
  {4}},\ \bibinfo {pages} {945} (\bibinfo {year} {2016})}\BibitemShut {NoStop}%
\bibitem [{\citenamefont {D'Errico}\ \emph {et~al.}(2014)\citenamefont
  {D'Errico}, \citenamefont {Lucioni}, \citenamefont {Tanzi}, \citenamefont
  {Gori}, \citenamefont {Roux}, \citenamefont {McCulloch}, \citenamefont
  {Giamarchi}, \citenamefont {Inguscio},\ and\ \citenamefont
  {Modugno}}]{DErrico-1D-Dis}%
  \BibitemOpen
  \bibfield  {author} {\bibinfo {author} {\bibfnamefont {C.}~\bibnamefont
  {D'Errico}}, \bibinfo {author} {\bibfnamefont {E.}~\bibnamefont {Lucioni}},
  \bibinfo {author} {\bibfnamefont {L.}~\bibnamefont {Tanzi}}, \bibinfo
  {author} {\bibfnamefont {L.}~\bibnamefont {Gori}}, \bibinfo {author}
  {\bibfnamefont {G.}~\bibnamefont {Roux}}, \bibinfo {author} {\bibfnamefont
  {I.~P.}\ \bibnamefont {McCulloch}}, \bibinfo {author} {\bibfnamefont
  {T.}~\bibnamefont {Giamarchi}}, \bibinfo {author} {\bibfnamefont
  {M.}~\bibnamefont {Inguscio}}, \ and\ \bibinfo {author} {\bibfnamefont
  {G.}~\bibnamefont {Modugno}},\ }\href {\doibase
  10.1103/PhysRevLett.113.095301} {\bibfield  {journal} {\bibinfo  {journal}
  {Phys. Rev. Lett.}\ }\textbf {\bibinfo {volume} {113}},\ \bibinfo {pages}
  {095301} (\bibinfo {year} {2014})}\BibitemShut {NoStop}%
\bibitem [{\citenamefont {Kondov}\ \emph {et~al.}(2015)\citenamefont {Kondov},
  \citenamefont {McGehee}, \citenamefont {Xu},\ and\ \citenamefont
  {DeMarco}}]{Kondov-MBL}%
  \BibitemOpen
  \bibfield  {author} {\bibinfo {author} {\bibfnamefont {S.~S.}\ \bibnamefont
  {Kondov}}, \bibinfo {author} {\bibfnamefont {W.~R.}\ \bibnamefont {McGehee}},
  \bibinfo {author} {\bibfnamefont {W.}~\bibnamefont {Xu}}, \ and\ \bibinfo
  {author} {\bibfnamefont {B.}~\bibnamefont {DeMarco}},\ }\href {\doibase
  10.1103/PhysRevLett.114.083002} {\bibfield  {journal} {\bibinfo  {journal}
  {Phys. Rev. Lett.}\ }\textbf {\bibinfo {volume} {114}},\ \bibinfo {pages}
  {083002} (\bibinfo {year} {2015})}\BibitemShut {NoStop}%
\bibitem [{\citenamefont {Schreiber}\ \emph {et~al.}(2015)\citenamefont
  {Schreiber}, \citenamefont {Hodgman}, \citenamefont {Bordia}, \citenamefont
  {L{\"u}schen}, \citenamefont {Fischer}, \citenamefont {Vosk}, \citenamefont
  {Altman}, \citenamefont {Schneider},\ and\ \citenamefont
  {Bloch}}]{Schreiber842}%
  \BibitemOpen
  \bibfield  {author} {\bibinfo {author} {\bibfnamefont {M.}~\bibnamefont
  {Schreiber}}, \bibinfo {author} {\bibfnamefont {S.~S.}\ \bibnamefont
  {Hodgman}}, \bibinfo {author} {\bibfnamefont {P.}~\bibnamefont {Bordia}},
  \bibinfo {author} {\bibfnamefont {H.~P.}\ \bibnamefont {L{\"u}schen}},
  \bibinfo {author} {\bibfnamefont {M.~H.}\ \bibnamefont {Fischer}}, \bibinfo
  {author} {\bibfnamefont {R.}~\bibnamefont {Vosk}}, \bibinfo {author}
  {\bibfnamefont {E.}~\bibnamefont {Altman}}, \bibinfo {author} {\bibfnamefont
  {U.}~\bibnamefont {Schneider}}, \ and\ \bibinfo {author} {\bibfnamefont
  {I.}~\bibnamefont {Bloch}},\ }\href {\doibase 10.1126/science.aaa7432}
  {\bibfield  {journal} {\bibinfo  {journal} {Science}\ }\textbf {\bibinfo
  {volume} {349}},\ \bibinfo {pages} {842} (\bibinfo {year}
  {2015})}\BibitemShut {NoStop}%
\bibitem [{\citenamefont {Choi}\ \emph {et~al.}(2016)\citenamefont {Choi},
  \citenamefont {Hild}, \citenamefont {Zeiher}, \citenamefont {Schau{\ss}},
  \citenamefont {Rubio-Abadal}, \citenamefont {Yefsah}, \citenamefont
  {Khemani}, \citenamefont {Huse}, \citenamefont {Bloch},\ and\ \citenamefont
  {Gross}}]{MBL-Gross}%
  \BibitemOpen
  \bibfield  {author} {\bibinfo {author} {\bibfnamefont {J.-y.}\ \bibnamefont
  {Choi}}, \bibinfo {author} {\bibfnamefont {S.}~\bibnamefont {Hild}}, \bibinfo
  {author} {\bibfnamefont {J.}~\bibnamefont {Zeiher}}, \bibinfo {author}
  {\bibfnamefont {P.}~\bibnamefont {Schau{\ss}}}, \bibinfo {author}
  {\bibfnamefont {A.}~\bibnamefont {Rubio-Abadal}}, \bibinfo {author}
  {\bibfnamefont {T.}~\bibnamefont {Yefsah}}, \bibinfo {author} {\bibfnamefont
  {V.}~\bibnamefont {Khemani}}, \bibinfo {author} {\bibfnamefont {D.~A.}\
  \bibnamefont {Huse}}, \bibinfo {author} {\bibfnamefont {I.}~\bibnamefont
  {Bloch}}, \ and\ \bibinfo {author} {\bibfnamefont {C.}~\bibnamefont
  {Gross}},\ }\href {\doibase 10.1126/science.aaf8834} {\bibfield  {journal}
  {\bibinfo  {journal} {Science}\ }\textbf {\bibinfo {volume} {352}},\ \bibinfo
  {pages} {1547} (\bibinfo {year} {2016})}\BibitemShut {NoStop}%
\bibitem [{\citenamefont {Yan}\ \emph {et~al.}(2016)\citenamefont {Yan},
  \citenamefont {Hui}, \citenamefont {Rigol},\ and\ \citenamefont
  {Scarola}}]{MBL-Vito}%
  \BibitemOpen
  \bibfield  {author} {\bibinfo {author} {\bibfnamefont {M.}~\bibnamefont
  {Yan}}, \bibinfo {author} {\bibfnamefont {H.-Y.}\ \bibnamefont {Hui}},
  \bibinfo {author} {\bibfnamefont {M.}~\bibnamefont {Rigol}}, \ and\ \bibinfo
  {author} {\bibfnamefont {V.~W.}\ \bibnamefont {Scarola}},\ }\href@noop {} {\
  (\bibinfo {year} {2016})},\ \Eprint {http://arxiv.org/abs/1606.03444}
  {arXiv:1606.03444} \BibitemShut {NoStop}%
\bibitem [{\citenamefont {Celi}\ \emph {et~al.}(2014)\citenamefont {Celi},
  \citenamefont {Massignan}, \citenamefont {Ruseckas}, \citenamefont {Goldman},
  \citenamefont {Spielman}, \citenamefont {Juzeli\ifmmode~\bar{u}\else
  \={u}\fi{}nas},\ and\ \citenamefont {Lewenstein}}]{Celi-ArtificialDim}%
  \BibitemOpen
  \bibfield  {author} {\bibinfo {author} {\bibfnamefont {A.}~\bibnamefont
  {Celi}}, \bibinfo {author} {\bibfnamefont {P.}~\bibnamefont {Massignan}},
  \bibinfo {author} {\bibfnamefont {J.}~\bibnamefont {Ruseckas}}, \bibinfo
  {author} {\bibfnamefont {N.}~\bibnamefont {Goldman}}, \bibinfo {author}
  {\bibfnamefont {I.~B.}\ \bibnamefont {Spielman}}, \bibinfo {author}
  {\bibfnamefont {G.}~\bibnamefont {Juzeli\ifmmode~\bar{u}\else
  \={u}\fi{}nas}}, \ and\ \bibinfo {author} {\bibfnamefont {M.}~\bibnamefont
  {Lewenstein}},\ }\href {\doibase 10.1103/PhysRevLett.112.043001} {\bibfield
  {journal} {\bibinfo  {journal} {Phys. Rev. Lett.}\ }\textbf {\bibinfo
  {volume} {112}},\ \bibinfo {pages} {043001} (\bibinfo {year}
  {2014})}\BibitemShut {NoStop}%
\bibitem [{\citenamefont {Stuhl}\ \emph {et~al.}(2015)\citenamefont {Stuhl},
  \citenamefont {Lu}, \citenamefont {Aycock}, \citenamefont {Genkina},\ and\
  \citenamefont {Spielman}}]{Stuhl-Edge-2015}%
  \BibitemOpen
  \bibfield  {author} {\bibinfo {author} {\bibfnamefont {B.~K.}\ \bibnamefont
  {Stuhl}}, \bibinfo {author} {\bibfnamefont {H.-I.}\ \bibnamefont {Lu}},
  \bibinfo {author} {\bibfnamefont {L.~M.}\ \bibnamefont {Aycock}}, \bibinfo
  {author} {\bibfnamefont {D.}~\bibnamefont {Genkina}}, \ and\ \bibinfo
  {author} {\bibfnamefont {I.~B.}\ \bibnamefont {Spielman}},\ }\href {\doibase
  10.1126/science.aaa8515} {\bibfield  {journal} {\bibinfo  {journal}
  {Science}\ }\textbf {\bibinfo {volume} {349}},\ \bibinfo {pages} {1514}
  (\bibinfo {year} {2015})}\BibitemShut {NoStop}%
\bibitem [{\citenamefont {Mancini}\ \emph {et~al.}(2015)\citenamefont
  {Mancini}, \citenamefont {Pagano}, \citenamefont {Cappellini}, \citenamefont
  {Livi}, \citenamefont {Rider}, \citenamefont {Catani}, \citenamefont {Sias},
  \citenamefont {Zoller}, \citenamefont {Inguscio}, \citenamefont {Dalmonte},\
  and\ \citenamefont {Fallani}}]{Fallani-chiral-2015}%
  \BibitemOpen
  \bibfield  {author} {\bibinfo {author} {\bibfnamefont {M.}~\bibnamefont
  {Mancini}}, \bibinfo {author} {\bibfnamefont {G.}~\bibnamefont {Pagano}},
  \bibinfo {author} {\bibfnamefont {G.}~\bibnamefont {Cappellini}}, \bibinfo
  {author} {\bibfnamefont {L.}~\bibnamefont {Livi}}, \bibinfo {author}
  {\bibfnamefont {M.}~\bibnamefont {Rider}}, \bibinfo {author} {\bibfnamefont
  {J.}~\bibnamefont {Catani}}, \bibinfo {author} {\bibfnamefont
  {C.}~\bibnamefont {Sias}}, \bibinfo {author} {\bibfnamefont {P.}~\bibnamefont
  {Zoller}}, \bibinfo {author} {\bibfnamefont {M.}~\bibnamefont {Inguscio}},
  \bibinfo {author} {\bibfnamefont {M.}~\bibnamefont {Dalmonte}}, \ and\
  \bibinfo {author} {\bibfnamefont {L.}~\bibnamefont {Fallani}},\ }\href
  {\doibase 10.1126/science.aaa8736} {\bibfield  {journal} {\bibinfo  {journal}
  {Science}\ }\textbf {\bibinfo {volume} {349}},\ \bibinfo {pages} {1510}
  (\bibinfo {year} {2015})}\BibitemShut {NoStop}%
\bibitem [{\citenamefont {Meier}\ \emph
  {et~al.}(2016{\natexlab{a}})\citenamefont {Meier}, \citenamefont {An},\ and\
  \citenamefont {Gadway}}]{Meier-AtomOptics}%
  \BibitemOpen
  \bibfield  {author} {\bibinfo {author} {\bibfnamefont {E.~J.}\ \bibnamefont
  {Meier}}, \bibinfo {author} {\bibfnamefont {F.~A.}\ \bibnamefont {An}}, \
  and\ \bibinfo {author} {\bibfnamefont {B.}~\bibnamefont {Gadway}},\ }\href
  {\doibase 10.1103/PhysRevA.93.051602} {\bibfield  {journal} {\bibinfo
  {journal} {Phys. Rev. A}\ }\textbf {\bibinfo {volume} {93}},\ \bibinfo
  {pages} {051602} (\bibinfo {year} {2016}{\natexlab{a}})}\BibitemShut
  {NoStop}%
\bibitem [{\citenamefont {Meier}\ \emph
  {et~al.}(2016{\natexlab{b}})\citenamefont {Meier}, \citenamefont {An},\ and\
  \citenamefont {Gadway}}]{Meier-SSH}%
  \BibitemOpen
  \bibfield  {author} {\bibinfo {author} {\bibfnamefont {E.~J.}\ \bibnamefont
  {Meier}}, \bibinfo {author} {\bibfnamefont {F.~A.}\ \bibnamefont {An}}, \
  and\ \bibinfo {author} {\bibfnamefont {B.}~\bibnamefont {Gadway}},\ }\href
  {http://dx.doi.org/10.1038/ncomms13986} {\bibfield  {journal} {\bibinfo
  {journal} {Nat. Commun.}\ }\textbf {\bibinfo {volume} {7}},\ \bibinfo {pages}
  {13986} (\bibinfo {year} {2016}{\natexlab{b}})}\BibitemShut {NoStop}%
\bibitem [{\citenamefont {An}\ \emph {et~al.}(2016)\citenamefont {An},
  \citenamefont {Meier},\ and\ \citenamefont {Gadway}}]{An-FluxLadder}%
  \BibitemOpen
  \bibfield  {author} {\bibinfo {author} {\bibfnamefont {F.~A.}\ \bibnamefont
  {An}}, \bibinfo {author} {\bibfnamefont {E.~J.}\ \bibnamefont {Meier}}, \
  and\ \bibinfo {author} {\bibfnamefont {B.}~\bibnamefont {Gadway}},\
  }\href@noop {} {\  (\bibinfo {year} {2016})},\ \Eprint
  {http://arxiv.org/abs/1609.09467} {arXiv:1609.09467} \BibitemShut {NoStop}%
\bibitem [{\citenamefont {Wall}\ \emph {et~al.}(2016)\citenamefont {Wall},
  \citenamefont {Koller}, \citenamefont {Li}, \citenamefont {Zhang},
  \citenamefont {Cooper}, \citenamefont {Ye},\ and\ \citenamefont
  {Rey}}]{Wall-Synth}%
  \BibitemOpen
  \bibfield  {author} {\bibinfo {author} {\bibfnamefont {M.~L.}\ \bibnamefont
  {Wall}}, \bibinfo {author} {\bibfnamefont {A.~P.}\ \bibnamefont {Koller}},
  \bibinfo {author} {\bibfnamefont {S.}~\bibnamefont {Li}}, \bibinfo {author}
  {\bibfnamefont {X.}~\bibnamefont {Zhang}}, \bibinfo {author} {\bibfnamefont
  {N.~R.}\ \bibnamefont {Cooper}}, \bibinfo {author} {\bibfnamefont
  {J.}~\bibnamefont {Ye}}, \ and\ \bibinfo {author} {\bibfnamefont {A.~M.}\
  \bibnamefont {Rey}},\ }\href {\doibase 10.1103/PhysRevLett.116.035301}
  {\bibfield  {journal} {\bibinfo  {journal} {Phys. Rev. Lett.}\ }\textbf
  {\bibinfo {volume} {116}},\ \bibinfo {pages} {035301} (\bibinfo {year}
  {2016})}\BibitemShut {NoStop}%
\bibitem [{\citenamefont {Kolkowitz}\ \emph {et~al.}(2016)\citenamefont
  {Kolkowitz}, \citenamefont {Bromley}, \citenamefont {Bothwell}, \citenamefont
  {Wall}, \citenamefont {Marti}, \citenamefont {Koller}, \citenamefont {Zhang},
  \citenamefont {Rey},\ and\ \citenamefont {Ye}}]{Kolkowitz-Synth}%
  \BibitemOpen
  \bibfield  {author} {\bibinfo {author} {\bibfnamefont {S.}~\bibnamefont
  {Kolkowitz}}, \bibinfo {author} {\bibfnamefont {S.~L.}\ \bibnamefont
  {Bromley}}, \bibinfo {author} {\bibfnamefont {T.}~\bibnamefont {Bothwell}},
  \bibinfo {author} {\bibfnamefont {M.~L.}\ \bibnamefont {Wall}}, \bibinfo
  {author} {\bibfnamefont {G.~E.}\ \bibnamefont {Marti}}, \bibinfo {author}
  {\bibfnamefont {A.~P.}\ \bibnamefont {Koller}}, \bibinfo {author}
  {\bibfnamefont {X.}~\bibnamefont {Zhang}}, \bibinfo {author} {\bibfnamefont
  {A.~M.}\ \bibnamefont {Rey}}, \ and\ \bibinfo {author} {\bibfnamefont
  {J.}~\bibnamefont {Ye}},\ }\href {http://dx.doi.org/10.1038/nature20811}
  {\bibfield  {journal} {\bibinfo  {journal} {Nature}\ }\textbf {\bibinfo
  {volume} {advance online publication}} (\bibinfo {year} {2016})}\BibitemShut
  {NoStop}%
\bibitem [{\citenamefont {Livi}\ \emph {et~al.}(2016)\citenamefont {Livi},
  \citenamefont {Cappellini}, \citenamefont {Diem}, \citenamefont {Franchi},
  \citenamefont {Clivati}, \citenamefont {Frittelli}, \citenamefont {Levi},
  \citenamefont {Calonico}, \citenamefont {Catani}, \citenamefont {Inguscio},\
  and\ \citenamefont {Fallani}}]{Livi-Synth}%
  \BibitemOpen
  \bibfield  {author} {\bibinfo {author} {\bibfnamefont {L.~F.}\ \bibnamefont
  {Livi}}, \bibinfo {author} {\bibfnamefont {G.}~\bibnamefont {Cappellini}},
  \bibinfo {author} {\bibfnamefont {M.}~\bibnamefont {Diem}}, \bibinfo {author}
  {\bibfnamefont {L.}~\bibnamefont {Franchi}}, \bibinfo {author} {\bibfnamefont
  {C.}~\bibnamefont {Clivati}}, \bibinfo {author} {\bibfnamefont
  {M.}~\bibnamefont {Frittelli}}, \bibinfo {author} {\bibfnamefont
  {F.}~\bibnamefont {Levi}}, \bibinfo {author} {\bibfnamefont {D.}~\bibnamefont
  {Calonico}}, \bibinfo {author} {\bibfnamefont {J.}~\bibnamefont {Catani}},
  \bibinfo {author} {\bibfnamefont {M.}~\bibnamefont {Inguscio}}, \ and\
  \bibinfo {author} {\bibfnamefont {L.}~\bibnamefont {Fallani}},\ }\href
  {\doibase 10.1103/PhysRevLett.117.220401} {\bibfield  {journal} {\bibinfo
  {journal} {Phys. Rev. Lett.}\ }\textbf {\bibinfo {volume} {117}},\ \bibinfo
  {pages} {220401} (\bibinfo {year} {2016})}\BibitemShut {NoStop}%
\bibitem [{\citenamefont {Gadway}(2015)}]{Gadway-KSPACE}%
  \BibitemOpen
  \bibfield  {author} {\bibinfo {author} {\bibfnamefont {B.}~\bibnamefont
  {Gadway}},\ }\href {\doibase 10.1103/PhysRevA.92.043606} {\bibfield
  {journal} {\bibinfo  {journal} {Phys. Rev. A}\ }\textbf {\bibinfo {volume}
  {92}},\ \bibinfo {pages} {043606} (\bibinfo {year} {2015})}\BibitemShut
  {NoStop}%
\bibitem [{\citenamefont {Amir}\ \emph {et~al.}(2009)\citenamefont {Amir},
  \citenamefont {Lahini},\ and\ \citenamefont {Perets}}]{LahiniDiffusion}%
  \BibitemOpen
  \bibfield  {author} {\bibinfo {author} {\bibfnamefont {A.}~\bibnamefont
  {Amir}}, \bibinfo {author} {\bibfnamefont {Y.}~\bibnamefont {Lahini}}, \ and\
  \bibinfo {author} {\bibfnamefont {H.~B.}\ \bibnamefont {Perets}},\ }\href
  {\doibase 10.1103/PhysRevE.79.050105} {\bibfield  {journal} {\bibinfo
  {journal} {Phys. Rev. E}\ }\textbf {\bibinfo {volume} {79}},\ \bibinfo
  {pages} {050105} (\bibinfo {year} {2009})}\BibitemShut {NoStop}%
\bibitem [{\citenamefont {Price}\ \emph {et~al.}(2016)\citenamefont {Price},
  \citenamefont {Ozawa},\ and\ \citenamefont {Goldman}}]{NateGold-TrapShake}%
  \BibitemOpen
  \bibfield  {author} {\bibinfo {author} {\bibfnamefont {H.~M.}\ \bibnamefont
  {Price}}, \bibinfo {author} {\bibfnamefont {T.}~\bibnamefont {Ozawa}}, \ and\
  \bibinfo {author} {\bibfnamefont {N.}~\bibnamefont {Goldman}},\ }\href@noop
  {} {\  (\bibinfo {year} {2016})},\ \Eprint {http://arxiv.org/abs/1605.09310}
  {arXiv:1605.09310} \BibitemShut {NoStop}%
\bibitem [{\citenamefont {Christodoulides}\ \emph {et~al.}(2003)\citenamefont
  {Christodoulides}, \citenamefont {Lederer},\ and\ \citenamefont
  {Silberberg}}]{Christ-NatRev}%
  \BibitemOpen
  \bibfield  {author} {\bibinfo {author} {\bibfnamefont {D.~N.}\ \bibnamefont
  {Christodoulides}}, \bibinfo {author} {\bibfnamefont {F.}~\bibnamefont
  {Lederer}}, \ and\ \bibinfo {author} {\bibfnamefont {Y.}~\bibnamefont
  {Silberberg}},\ }\href {\doibase 10.1038/nature01936} {\bibfield  {journal}
  {\bibinfo  {journal} {Nature}\ }\textbf {\bibinfo {volume} {424}},\ \bibinfo
  {pages} {817} (\bibinfo {year} {2003})}\BibitemShut {NoStop}%
\bibitem [{\citenamefont {Schwartz}\ \emph {et~al.}(2007)\citenamefont
  {Schwartz}, \citenamefont {Bartal}, \citenamefont {Fishman},\ and\
  \citenamefont {Segev}}]{And-Light-Seg-07}%
  \BibitemOpen
  \bibfield  {author} {\bibinfo {author} {\bibfnamefont {T.}~\bibnamefont
  {Schwartz}}, \bibinfo {author} {\bibfnamefont {G.}~\bibnamefont {Bartal}},
  \bibinfo {author} {\bibfnamefont {S.}~\bibnamefont {Fishman}}, \ and\
  \bibinfo {author} {\bibfnamefont {M.}~\bibnamefont {Segev}},\ }\href
  {\doibase 10.1038/nature05623} {\bibfield  {journal} {\bibinfo  {journal}
  {Nature}\ }\textbf {\bibinfo {volume} {446}},\ \bibinfo {pages} {52}
  (\bibinfo {year} {2007})}\BibitemShut {NoStop}%
\bibitem [{\citenamefont {Szameit}\ and\ \citenamefont
  {Nolte}(2010)}]{SzameitReview-2010}%
  \BibitemOpen
  \bibfield  {author} {\bibinfo {author} {\bibfnamefont {A.}~\bibnamefont
  {Szameit}}\ and\ \bibinfo {author} {\bibfnamefont {S.}~\bibnamefont
  {Nolte}},\ }\href {\doibase 10.1088/0953-4075/43/16/163001} {\bibfield
  {journal} {\bibinfo  {journal} {J. Phys. B}\ }\textbf {\bibinfo {volume}
  {43}},\ \bibinfo {pages} {163001} (\bibinfo {year} {2010})}\BibitemShut
  {NoStop}%
\bibitem [{\citenamefont {Segev}\ \emph {et~al.}(2013)\citenamefont {Segev},
  \citenamefont {Silberberg},\ and\ \citenamefont
  {Christodoulides}}]{AndersonLight-Review}%
  \BibitemOpen
  \bibfield  {author} {\bibinfo {author} {\bibfnamefont {M.}~\bibnamefont
  {Segev}}, \bibinfo {author} {\bibfnamefont {Y.}~\bibnamefont {Silberberg}}, \
  and\ \bibinfo {author} {\bibfnamefont {D.~N.}\ \bibnamefont
  {Christodoulides}},\ }\href {\doibase 10.1038/nphoton.2013.30} {\bibfield
  {journal} {\bibinfo  {journal} {Nat. Photon.}\ }\textbf {\bibinfo {volume}
  {7}},\ \bibinfo {pages} {197} (\bibinfo {year} {2013})}\BibitemShut {NoStop}%
\bibitem [{\citenamefont {Aspuru-Guzik}\ and\ \citenamefont
  {Walther}(2012)}]{PhotRev-NatPhys-2012}%
  \BibitemOpen
  \bibfield  {author} {\bibinfo {author} {\bibfnamefont {A.}~\bibnamefont
  {Aspuru-Guzik}}\ and\ \bibinfo {author} {\bibfnamefont {P.}~\bibnamefont
  {Walther}},\ }\href {\doibase doi:10.1038/nphys2253} {\bibfield  {journal}
  {\bibinfo  {journal} {Nat. Phys.}\ }\textbf {\bibinfo {volume} {8}},\
  \bibinfo {pages} {285} (\bibinfo {year} {2012})}\BibitemShut {NoStop}%
\bibitem [{\citenamefont {Lee}\ and\ \citenamefont
  {Fisher}(1981)}]{Lee-Fisher-RandomFlux-1981}%
  \BibitemOpen
  \bibfield  {author} {\bibinfo {author} {\bibfnamefont {P.~A.}\ \bibnamefont
  {Lee}}\ and\ \bibinfo {author} {\bibfnamefont {D.~S.}\ \bibnamefont
  {Fisher}},\ }\href {\doibase 10.1103/PhysRevLett.47.882} {\bibfield
  {journal} {\bibinfo  {journal} {Phys. Rev. Lett.}\ }\textbf {\bibinfo
  {volume} {47}},\ \bibinfo {pages} {882} (\bibinfo {year} {1981})}\BibitemShut
  {NoStop}%
\bibitem [{\citenamefont {Ludwig}\ \emph {et~al.}(1994)\citenamefont {Ludwig},
  \citenamefont {Fisher}, \citenamefont {Shankar},\ and\ \citenamefont
  {Grinstein}}]{RandomField1}%
  \BibitemOpen
  \bibfield  {author} {\bibinfo {author} {\bibfnamefont {A.~W.~W.}\
  \bibnamefont {Ludwig}}, \bibinfo {author} {\bibfnamefont {M.~P.~A.}\
  \bibnamefont {Fisher}}, \bibinfo {author} {\bibfnamefont {R.}~\bibnamefont
  {Shankar}}, \ and\ \bibinfo {author} {\bibfnamefont {G.}~\bibnamefont
  {Grinstein}},\ }\href {\doibase 10.1103/PhysRevB.50.7526} {\bibfield
  {journal} {\bibinfo  {journal} {Phys. Rev. B}\ }\textbf {\bibinfo {volume}
  {50}},\ \bibinfo {pages} {7526} (\bibinfo {year} {1994})}\BibitemShut
  {NoStop}%
\bibitem [{\citenamefont {Chamon}\ \emph {et~al.}(1996)\citenamefont {Chamon},
  \citenamefont {Mudry},\ and\ \citenamefont {Wen}}]{RandomField2}%
  \BibitemOpen
  \bibfield  {author} {\bibinfo {author} {\bibfnamefont {C.~d.~C.}\
  \bibnamefont {Chamon}}, \bibinfo {author} {\bibfnamefont {C.}~\bibnamefont
  {Mudry}}, \ and\ \bibinfo {author} {\bibfnamefont {X.-G.}\ \bibnamefont
  {Wen}},\ }\href {\doibase 10.1103/PhysRevLett.77.4194} {\bibfield  {journal}
  {\bibinfo  {journal} {Phys. Rev. Lett.}\ }\textbf {\bibinfo {volume} {77}},\
  \bibinfo {pages} {4194} (\bibinfo {year} {1996})}\BibitemShut {NoStop}%
\bibitem [{\citenamefont {Brun}\ \emph {et~al.}(2003)\citenamefont {Brun},
  \citenamefont {Carteret},\ and\ \citenamefont
  {Ambainis}}]{QtoC-Theory-Decoh}%
  \BibitemOpen
  \bibfield  {author} {\bibinfo {author} {\bibfnamefont {T.~A.}\ \bibnamefont
  {Brun}}, \bibinfo {author} {\bibfnamefont {H.~A.}\ \bibnamefont {Carteret}},
  \ and\ \bibinfo {author} {\bibfnamefont {A.}~\bibnamefont {Ambainis}},\
  }\href {\doibase 10.1103/PhysRevLett.91.130602} {\bibfield  {journal}
  {\bibinfo  {journal} {Phys. Rev. Lett.}\ }\textbf {\bibinfo {volume} {91}},\
  \bibinfo {pages} {130602} (\bibinfo {year} {2003})}\BibitemShut {NoStop}%
\bibitem [{\citenamefont {Broome}\ \emph {et~al.}(2010)\citenamefont {Broome},
  \citenamefont {Fedrizzi}, \citenamefont {Lanyon}, \citenamefont {Kassal},
  \citenamefont {Aspuru-Guzik},\ and\ \citenamefont
  {White}}]{Broome-White-QtoC-Decoherence}%
  \BibitemOpen
  \bibfield  {author} {\bibinfo {author} {\bibfnamefont {M.~A.}\ \bibnamefont
  {Broome}}, \bibinfo {author} {\bibfnamefont {A.}~\bibnamefont {Fedrizzi}},
  \bibinfo {author} {\bibfnamefont {B.~P.}\ \bibnamefont {Lanyon}}, \bibinfo
  {author} {\bibfnamefont {I.}~\bibnamefont {Kassal}}, \bibinfo {author}
  {\bibfnamefont {A.}~\bibnamefont {Aspuru-Guzik}}, \ and\ \bibinfo {author}
  {\bibfnamefont {A.~G.}\ \bibnamefont {White}},\ }\href {\doibase
  10.1103/PhysRevLett.104.153602} {\bibfield  {journal} {\bibinfo  {journal}
  {Phys. Rev. Lett.}\ }\textbf {\bibinfo {volume} {104}},\ \bibinfo {pages}
  {153602} (\bibinfo {year} {2010})}\BibitemShut {NoStop}%
\bibitem [{\citenamefont {Schreiber}\ \emph {et~al.}(2011)\citenamefont
  {Schreiber}, \citenamefont {Cassemiro}, \citenamefont {Poto\v{c}ek},
  \citenamefont {G\'{a}bris}, \citenamefont {Jex},\ and\ \citenamefont
  {Silberhorn}}]{Silberhorn-DisorderAndDecoherence}%
  \BibitemOpen
  \bibfield  {author} {\bibinfo {author} {\bibfnamefont {A.}~\bibnamefont
  {Schreiber}}, \bibinfo {author} {\bibfnamefont {K.~N.}\ \bibnamefont
  {Cassemiro}}, \bibinfo {author} {\bibfnamefont {V.}~\bibnamefont
  {Poto\v{c}ek}}, \bibinfo {author} {\bibfnamefont {A.}~\bibnamefont
  {G\'{a}bris}}, \bibinfo {author} {\bibfnamefont {I.}~\bibnamefont {Jex}}, \
  and\ \bibinfo {author} {\bibfnamefont {C.}~\bibnamefont {Silberhorn}},\
  }\href {\doibase 10.1103/PhysRevLett.106.180403} {\bibfield  {journal}
  {\bibinfo  {journal} {Phys. Rev. Lett.}\ }\textbf {\bibinfo {volume} {106}},\
  \bibinfo {pages} {180403} (\bibinfo {year} {2011})}\BibitemShut {NoStop}%
\bibitem [{\citenamefont {Karski}\ \emph {et~al.}(2009)\citenamefont {Karski},
  \citenamefont {F{\"o}rster}, \citenamefont {Choi}, \citenamefont {Steffen},
  \citenamefont {Alt}, \citenamefont {Meschede},\ and\ \citenamefont
  {Widera}}]{Karski174}%
  \BibitemOpen
  \bibfield  {author} {\bibinfo {author} {\bibfnamefont {M.}~\bibnamefont
  {Karski}}, \bibinfo {author} {\bibfnamefont {L.}~\bibnamefont {F{\"o}rster}},
  \bibinfo {author} {\bibfnamefont {J.-M.}\ \bibnamefont {Choi}}, \bibinfo
  {author} {\bibfnamefont {A.}~\bibnamefont {Steffen}}, \bibinfo {author}
  {\bibfnamefont {W.}~\bibnamefont {Alt}}, \bibinfo {author} {\bibfnamefont
  {D.}~\bibnamefont {Meschede}}, \ and\ \bibinfo {author} {\bibfnamefont
  {A.}~\bibnamefont {Widera}},\ }\href {\doibase 10.1126/science.1174436}
  {\bibfield  {journal} {\bibinfo  {journal} {Science}\ }\textbf {\bibinfo
  {volume} {325}},\ \bibinfo {pages} {174} (\bibinfo {year}
  {2009})}\BibitemShut {NoStop}%
\bibitem [{\citenamefont {Fukuhara}\ \emph {et~al.}(2013)\citenamefont
  {Fukuhara}, \citenamefont {Kantian}, \citenamefont {Endres}, \citenamefont
  {Cheneau}, \citenamefont {Schau{\ss}}, \citenamefont {Hild}, \citenamefont
  {Bellem}, \citenamefont {Schollw{\"o}ck}, \citenamefont {Giamarchi},
  \citenamefont {Gross}, \citenamefont {Bloch},\ and\ \citenamefont
  {Kuhr}}]{fukuhara:quantum_2013}%
  \BibitemOpen
  \bibfield  {author} {\bibinfo {author} {\bibfnamefont {T.}~\bibnamefont
  {Fukuhara}}, \bibinfo {author} {\bibfnamefont {A.}~\bibnamefont {Kantian}},
  \bibinfo {author} {\bibfnamefont {M.}~\bibnamefont {Endres}}, \bibinfo
  {author} {\bibfnamefont {M.}~\bibnamefont {Cheneau}}, \bibinfo {author}
  {\bibfnamefont {P.}~\bibnamefont {Schau{\ss}}}, \bibinfo {author}
  {\bibfnamefont {S.}~\bibnamefont {Hild}}, \bibinfo {author} {\bibfnamefont
  {D.}~\bibnamefont {Bellem}}, \bibinfo {author} {\bibfnamefont
  {U.}~\bibnamefont {Schollw{\"o}ck}}, \bibinfo {author} {\bibfnamefont
  {T.}~\bibnamefont {Giamarchi}}, \bibinfo {author} {\bibfnamefont
  {C.}~\bibnamefont {Gross}}, \bibinfo {author} {\bibfnamefont
  {I.}~\bibnamefont {Bloch}}, \ and\ \bibinfo {author} {\bibfnamefont
  {S.}~\bibnamefont {Kuhr}},\ }\href {\doibase 10.1038/nphys2561} {\bibfield
  {journal} {\bibinfo  {journal} {Nat. Phys.}\ }\textbf {\bibinfo {volume}
  {9}},\ \bibinfo {pages} {235} (\bibinfo {year} {2013})}\BibitemShut {NoStop}%
\bibitem [{\citenamefont {Osterloh}\ \emph {et~al.}(2005)\citenamefont
  {Osterloh}, \citenamefont {Baig}, \citenamefont {Santos}, \citenamefont
  {Zoller},\ and\ \citenamefont {Lewenstein}}]{Osterloh-NonAbel-2005PRL}%
  \BibitemOpen
  \bibfield  {author} {\bibinfo {author} {\bibfnamefont {K.}~\bibnamefont
  {Osterloh}}, \bibinfo {author} {\bibfnamefont {M.}~\bibnamefont {Baig}},
  \bibinfo {author} {\bibfnamefont {L.}~\bibnamefont {Santos}}, \bibinfo
  {author} {\bibfnamefont {P.}~\bibnamefont {Zoller}}, \ and\ \bibinfo {author}
  {\bibfnamefont {M.}~\bibnamefont {Lewenstein}},\ }\href {\doibase
  10.1103/PhysRevLett.95.010403} {\bibfield  {journal} {\bibinfo  {journal}
  {Phys. Rev. Lett.}\ }\textbf {\bibinfo {volume} {95}},\ \bibinfo {pages}
  {010403} (\bibinfo {year} {2005})}\BibitemShut {NoStop}%
\bibitem [{\citenamefont {Rudner}\ \emph {et~al.}(2013)\citenamefont {Rudner},
  \citenamefont {Lindner}, \citenamefont {Berg},\ and\ \citenamefont
  {Levin}}]{Rudner}%
  \BibitemOpen
  \bibfield  {author} {\bibinfo {author} {\bibfnamefont {M.~S.}\ \bibnamefont
  {Rudner}}, \bibinfo {author} {\bibfnamefont {N.~H.}\ \bibnamefont {Lindner}},
  \bibinfo {author} {\bibfnamefont {E.}~\bibnamefont {Berg}}, \ and\ \bibinfo
  {author} {\bibfnamefont {M.}~\bibnamefont {Levin}},\ }\href {\doibase
  10.1103/PhysRevX.3.031005} {\bibfield  {journal} {\bibinfo  {journal} {Phys.
  Rev. X}\ }\textbf {\bibinfo {volume} {3}},\ \bibinfo {pages} {031005}
  (\bibinfo {year} {2013})}\BibitemShut {NoStop}%
\bibitem [{\citenamefont {Mukherjee}\ \emph {et~al.}(2017)\citenamefont
  {Mukherjee}, \citenamefont {Spracklen}, \citenamefont {Valiente},
  \citenamefont {Andersson}, \citenamefont {\"{O}hberg}, \citenamefont
  {Goldman},\ and\ \citenamefont {Thomson}}]{Goldman-Floq-And}%
  \BibitemOpen
  \bibfield  {author} {\bibinfo {author} {\bibfnamefont {S.}~\bibnamefont
  {Mukherjee}}, \bibinfo {author} {\bibfnamefont {A.}~\bibnamefont
  {Spracklen}}, \bibinfo {author} {\bibfnamefont {M.}~\bibnamefont {Valiente}},
  \bibinfo {author} {\bibfnamefont {E.}~\bibnamefont {Andersson}}, \bibinfo
  {author} {\bibfnamefont {P.}~\bibnamefont {\"{O}hberg}}, \bibinfo {author}
  {\bibfnamefont {N.}~\bibnamefont {Goldman}}, \ and\ \bibinfo {author}
  {\bibfnamefont {R.~R.}\ \bibnamefont {Thomson}},\ }\href {\doibase
  10.1038/ncomms13918} {\bibfield  {journal} {\bibinfo  {journal} {Nat.
  Commun.}\ }\textbf {\bibinfo {volume} {8}},\ \bibinfo {pages} {13918}
  (\bibinfo {year} {2017})}\BibitemShut {NoStop}%
\bibitem [{\citenamefont {Maczewsky}\ \emph {et~al.}(2017)\citenamefont
  {Maczewsky}, \citenamefont {Zeuner}, \citenamefont {Nolte},\ and\
  \citenamefont {Szameit}}]{Szameit-Floq-And}%
  \BibitemOpen
  \bibfield  {author} {\bibinfo {author} {\bibfnamefont {L.~J.}\ \bibnamefont
  {Maczewsky}}, \bibinfo {author} {\bibfnamefont {J.~M.}\ \bibnamefont
  {Zeuner}}, \bibinfo {author} {\bibfnamefont {S.}~\bibnamefont {Nolte}}, \
  and\ \bibinfo {author} {\bibfnamefont {A.}~\bibnamefont {Szameit}},\ }\href
  {\doibase 10.1038/ncomms13756} {\bibfield  {journal} {\bibinfo  {journal}
  {Nat. Commun.}\ }\textbf {\bibinfo {volume} {8}},\ \bibinfo {pages} {13756}
  (\bibinfo {year} {2017})}\BibitemShut {NoStop}%
\bibitem [{\citenamefont {Titum}\ \emph {et~al.}(2016)\citenamefont {Titum},
  \citenamefont {Berg}, \citenamefont {Rudner}, \citenamefont {Refael},\ and\
  \citenamefont {Lindner}}]{Titum-Floq-And}%
  \BibitemOpen
  \bibfield  {author} {\bibinfo {author} {\bibfnamefont {P.}~\bibnamefont
  {Titum}}, \bibinfo {author} {\bibfnamefont {E.}~\bibnamefont {Berg}},
  \bibinfo {author} {\bibfnamefont {M.~S.}\ \bibnamefont {Rudner}}, \bibinfo
  {author} {\bibfnamefont {G.}~\bibnamefont {Refael}}, \ and\ \bibinfo {author}
  {\bibfnamefont {N.~H.}\ \bibnamefont {Lindner}},\ }\href {\doibase
  10.1103/PhysRevX.6.021013} {\bibfield  {journal} {\bibinfo  {journal} {Phys.
  Rev. X}\ }\textbf {\bibinfo {volume} {6}},\ \bibinfo {pages} {021013}
  (\bibinfo {year} {2016})}\BibitemShut {NoStop}%
\bibitem [{\citenamefont {Kosior}\ and\ \citenamefont
  {Sacha}(2017)}]{Kosior-RandomFractals}%
  \BibitemOpen
  \bibfield  {author} {\bibinfo {author} {\bibfnamefont {A.}~\bibnamefont
  {Kosior}}\ and\ \bibinfo {author} {\bibfnamefont {K.}~\bibnamefont {Sacha}},\
  }\href@noop {} {\  (\bibinfo {year} {2017})},\ \Eprint
  {http://arxiv.org/abs/1701.04274} {arXiv:1701.04274} \BibitemShut {NoStop}%
\bibitem [{\citenamefont {Dunlap}\ \emph {et~al.}(1990)\citenamefont {Dunlap},
  \citenamefont {Wu},\ and\ \citenamefont {Phillips}}]{Dunlap-1990}%
  \BibitemOpen
  \bibfield  {author} {\bibinfo {author} {\bibfnamefont {D.~H.}\ \bibnamefont
  {Dunlap}}, \bibinfo {author} {\bibfnamefont {H.-L.}\ \bibnamefont {Wu}}, \
  and\ \bibinfo {author} {\bibfnamefont {P.~W.}\ \bibnamefont {Phillips}},\
  }\href {\doibase 10.1103/PhysRevLett.65.88} {\bibfield  {journal} {\bibinfo
  {journal} {Phys. Rev. Lett.}\ }\textbf {\bibinfo {volume} {65}},\ \bibinfo
  {pages} {88} (\bibinfo {year} {1990})}\BibitemShut {NoStop}%
\bibitem [{\citenamefont {Rolston}\ and\ \citenamefont
  {Phillips}(2002)}]{Rolston-NL-2002}%
  \BibitemOpen
  \bibfield  {author} {\bibinfo {author} {\bibfnamefont {S.~L.}\ \bibnamefont
  {Rolston}}\ and\ \bibinfo {author} {\bibfnamefont {W.~D.}\ \bibnamefont
  {Phillips}},\ }\href {\doibase 10.1038/416219a} {\bibfield  {journal}
  {\bibinfo  {journal} {Nature}\ }\textbf {\bibinfo {volume} {416}},\ \bibinfo
  {pages} {219} (\bibinfo {year} {2002})}\BibitemShut {NoStop}%
\bibitem [{\citenamefont {Aleiner}\ \emph {et~al.}(2010)\citenamefont
  {Aleiner}, \citenamefont {Altshuler},\ and\ \citenamefont
  {Shlyapnikov}}]{aleiner:finite_temperature_disorder_2010}%
  \BibitemOpen
  \bibfield  {author} {\bibinfo {author} {\bibfnamefont {I.~L.}\ \bibnamefont
  {Aleiner}}, \bibinfo {author} {\bibfnamefont {B.~L.}\ \bibnamefont
  {Altshuler}}, \ and\ \bibinfo {author} {\bibfnamefont {G.~V.}\ \bibnamefont
  {Shlyapnikov}},\ }\href {\doibase 10.1038/nphys1758} {\bibfield  {journal}
  {\bibinfo  {journal} {Nat. Phys.}\ }\textbf {\bibinfo {volume} {6}},\
  \bibinfo {pages} {900} (\bibinfo {year} {2010})}\BibitemShut {NoStop}%
\end{thebibliography}
\end{document}